\documentclass[prb,twocolumn,showpacs,preprintnumbers,amsmath,amssymb]{revtex4}
\usepackage{amsfonts}
\usepackage{graphicx}
\usepackage{dcolumn}
\usepackage{float}
\usepackage{bm}

\newcommand{\com}[1]{}

\newcommand{\bnu}{\boldsymbol{\nu}}

\newcommand{\bS}{\mathbf{S}}
\newcommand{\q}{\mathbf{q}}
\newcommand{\p}{\mathbf{p}}

\newcommand{\h}{\mathbf{h}}

\newcommand{\K}{\mathbf{K}}
\newcommand{\rr}{\mathbf{r}}

\newcommand{\bsigma}{{\boldsymbol\sigma}}
\newcommand{\x}{{\hat x}}
\newcommand{\y}{{\hat y}}
\newcommand{\z}{{\hat z}}
\newcommand{\up}{{\uparrow}}
\newcommand{\dn}{{\downarrow}}
\newcommand{\ve}{{\varepsilon}}

\newcommand{\beq}{\begin{equation}}
\newcommand{\eeq}{\end{equation}}
\newcommand{\beqa}{\begin{eqnarray}}
\newcommand{\eeqa}{\end{eqnarray}}


\begin{document}

\title{Majorana fermions in superconducting helical magnets}

\author{Ivar Martin}
\affiliation{Theoretical Division, Los Alamos National Laboratory, Los Alamos, NM
87545, USA}

\author{Alberto F. Morpurgo}
\affiliation{DPMC and GAP, Universit\'e de Gen\`eve, CH-1211 Gen\`eve 4, Switzerland}

\begin{abstract}

In a variety of rare-earth based compounds singlet superconductivity coexists with helical magnetism. Here we demonstrate that surfaces of these systems should generically host a finite density of zero-energy Majorana modes. In the limit of vanishing disorder, these modes lead to a divergent contribution to zero-energy density of states and to zero-temperature entropy proportional to the sample surface area.  When confined to a wire geometry, a discrete number of Majorana modes can be isolated. The relatively large characteristic energy scales for superconductivity and magnetism, compared to other proposals, as well as the lack of need for fine-tuning, make helical magnetic superconducting compounds favorable for the observation and experimental investigation of Majorana fermions.

\end{abstract}


\maketitle

\section{Introduction}
Majorana fermions (MF) have attracted considerable attention recently as a promising tool for topological quantum computing, quantum information storage \cite{Kitaev, Nayak RMP, Ivanov}, and for unconventional quantum transport phemonena \cite{been}.  The appeal of MF is in their non-local character, when realized in topological states of matter: a single binary quantum degree of freedom can be split into two MF spatially isolated from each other, and thus protected from decoherence caused by local environmental perturbations. An active search for material realizations of MF is under way. A list of the candidate systems includes chiral p-wave superconductors \cite{Read-Green, Ivanov}, topological insulator with proximity induced superconductivity \cite{Fu-Kane}, noncentrosymmetric superconductors \cite{sato-noncentro, sau-semi, Potter}, and 5/2 Quantum Hall state \cite{Moore-Read}. In these systems Majorana zero modes can appear in the vortex cores or at the edges \cite{kitaev1d, von Oppen, das Sarma}.

The practical implementation of most of these proposals is hampered by the small energy scale which separates the Majorana modes from the electronic continuum, and by the need to precisely tune different system parameters. For instance, recently, an elegant proposal to create and manipulate Majorana fermions  was made based on superconducting nano-wires in the presence of Rashba spin-orbit interaction and magnetic field \cite{von Oppen, das Sarma, Alicea}. It requires a delicate balancing of the various coupling constants: a gap in the electronic spectrum, which has to be sufficiently large to prevent detrimental effects of disorder, has to be opened by an external magnetic field, which, at the same time, cannot be too large to avoid destroying superconductivity. It is clear that progress in the experimental search of Majorana fermions would be greatly facilitated by the identification of candidate systems, where Majorana physics can be explored in a larger energy range, without the need to satisfy highly demanding experimental constraints.

Here, we propose the use of helical magnetic superconductors (HMS) as such candidate system. While the bulk properties of HMS have been investigated in the past, the sub-gap surface states in this class of materials have remained unexplored. We show that the spin-helix in these systems plays a role analogous to the combination of the  spin-orbit interaction and of the magnetic field in the case of superconducting Rashba nanowires, and leads to formation of zero-energy Majorana surface states. Importantly,  however, the pair-breaking effects of the exchange interaction in HMS are dramatically reduced due to the spatial modulation of magnetism, as is evidenced by a large variety of existing materials  \cite{suhl-anderson, bulaevskii-kulic, buzdin-kulic, sigrist} that exhibit coexisting helical magnetism and superconductivity with relatively large transition temperatures (e.g, in ErNi$_2$B$_2$C the superconducting critical temperature is 10.5 K and the Neel temperature is 6.8 K). This property makes the readily available HMS a favorable systems for the detection of Majorana zero modes, without the need to match demanding experimental constraints, and offering an increase of an order of magnitude -- or more -- in the energy scale relevant for Majorana physics as compared to a typical semiconductor based system. While in idealized truly 1D systems the HMS and the superconducting Rashba systems are unitarily equivalent, in wires of finite width or in bulk systems -- relevant to actual experiments -- they are not. As we demonstrate, a remarkable feature of  2D and 3D HMS is that they generically possess a finite density of zero-energy Majorana surface modes. The resulting
large density of states can be accessed with surface probes (tunneling microscopy or photoemission) and should also manifest in thermodynamical measurements as a residual low-temperature entropy proportional to the surface area of the sample. 

The interplay of superconductivity and magnetism has long been a subject of active research. Uniform ferromagnetism is known to be strongly antagonistic to singlet superconductivity, due to the orbital magnetic field, which disrupts the superconducting phase by introducing vortices, and the exchange field that tends to split the singlet Cooper pairs. However, if the magnetization is spatially non-uniform on a scale smaller than the superconducting coherence length, both of these effects are dramatically suppressed. In fact, the onset of superconductivity itself can drive a uniform ferromagnet into a non-uniform (helical, or Ising-domain) state \cite{suhl-anderson, bulaevskii-kulic, buzdin-kulic, sigrist}. Examples of materials where helical magnetism coexists with superconductivity are those where magnetism originates from the partially filled f-orbitals of rare earth atoms, as is the case for the compounds such as HoMo$_6$S$_8$, ErRh$_4$B$_4$, TmNi$_2$B$_2$C \cite{buzdin-kulic}.  Because the f-orbitals are very compact, they directly interact only with the itinerant electrons, which can derive from the higher energy delocalized bands of the same atoms, or from other elements in the compound. The exchange interaction between local moments and itinerant electrons is typically much smaller than the itinerant electron bandwidth. Under these conditions, the local moments tend to order into a helical state, whose optimal pitch $K$ is determined by the maximum of the itinerant electron spin susceptibility $\chi(\q)$ (the Ruderman-Kittel-Kasuya-Yosida mechanism, \cite{RKKY}). As long as the superconducting coherence length $\xi$ is longer than the period of the helix, $K\xi \gg 1$, superconductivity is only weakly affected.

\section{Model}

Our analysis is based upon the mean-field superconducting Kondo lattice model,
\beqa
H &=&  -\sum_{i,j}{t_{ij} c_{i\alpha}^\dag c_{j\alpha}} -\sum_i\mu c_{i\alpha}^\dag c_{i\alpha}+ J   c_{i\alpha}^\dag \bS_i\cdot\bsigma_{\alpha\beta} c_{i\beta}\nonumber\\
&&+\Delta c_{i\up}^\dag c_{i\dn}^\dag + \Delta^* c_{i\dn} c_{i\up} \label{eq:H}
\eeqa
which has been applied with success for theoretical description of the bulk properties of HMS \cite{bulaevskii-kulic, buzdin-kulic}. This Hamiltonian describes electrons hopping between lattice  sites $i,j$ and interacting with classical magenetic order parameter $\bS_i$. Here $t_{i,j}$ is the intersite hopping, $J$ is the exchange interaction constant,  $\bsigma = (\sigma^x, \sigma^y, \sigma^z)$ is the vector of Pauli matrices, and $c_{i\alpha}$ is the operator of electron annihilation on site $i$ with spin $\alpha$.  For the helical state we choose $\bS (\rr) = (\cos Kz, \sin Kz, 0)$.  Even in the presence of helical magnetism the paring amplitude $\Delta$ in the singlet channel remains essentially uniform \cite{bulaevskii-kulic, buzdin-kulic}.

The Hamiltonian (\ref{eq:H}) is position dependent; however, the explicit $z$-dependence can be eliminated by performing a gauge transformation on electrons, $c_{i\up} \rightarrow c_{i\up} e^{iKz/2}$ and $c_{i\dn} \rightarrow c_{i\dn} e^{-iKz/2}$, which leaves the superconducting term invariant. This transformation changes the kinetic energy, which in momentum space becomes
\beqa
\ve_p c_{\p\alpha}^\dag c_{\p\alpha} \rightarrow \tilde\ve(\p) c_{\p\alpha}^\dag c_{\p\alpha} + h_z(\p) c_{\p\alpha}^\dag \sigma^z_{\alpha\beta}c_{\p\beta}.\nonumber
\eeqa
Here $\tilde\ve(\p)= (\ve_{\p-\K/2} + \ve_{\p+\K/2})/2$ and $h_z(\p)= (\ve_{\p+\K/2} - \ve_{\p-\K/2})/2$. If we assume here that the crystal structure is centrosymmetric; then $\tilde\ve$ is symmetric with respect to all components of $\p$ and $\K$, while $h$ is antisymmetric with respect to $p_z$ and $K$, and symmetric with respect to $p_{x,y}$. The complete Hamiltonian is

\beqa
H &=&  \sum_\p[\tilde\ve(\p) - \mu] c_{\p\alpha}^\dag c_{\p\alpha} \nonumber \\
&&- h_z(\p) c_{\p\alpha}^\dag \sigma^z_{\alpha\beta}c_{\p\beta}- J   c_{\p\alpha}^\dag \sigma^x_{\alpha\beta} c_{\p\beta}\nonumber\\
&&-\Delta c_{\p\up}^\dag c_{-\p\dn}^\dag - \Delta^* c_{-\p\dn} c_{\p\up}. \label{eq:H2}
\eeqa
It describes superconducting electrons in a momentum-dependent exchange field, $\h = (J,0, h_z)$. More specifically, the normal part is equivalent to an equal mixture of the Rashba and the Dresselhaus spin-orbit interactions, while $J$ plays a role of an external Zeeman field. For $J = 0$ there are two Fermi surfaces intersecting at $p_z = 0$ (where $h_z = 0$). Finite $J$ splits this degeneracy. The local spin quantization axis is rotated by angle $\phi(\p) = \tan^{-1}[J/h_z(\p)]$ relative to the $\z$ axis in the $\x$-$\z$ plane.

It is useful to transform the Hamiltonian to the basis that is locally aligned with the momentum-dependent field $\h$,
\beqa
\left(\begin{array}{c} c_{\p\up}\\c_{\p\dn}\end{array}\right) = e^{-i\sigma^y \phi(\p)/2} \left(\begin{array}{c} d_{\p\up}\\d_{\p\dn}\end{array}\right).
\eeqa
In this basis the Hamiltonian becomes,
\beqa
H &=&  \sum_\p[\tilde\ve(\p) - |\h(\p)|\sigma^z_{\alpha\beta} - \mu] d_{\p\alpha}^\dag d_{\p\beta}  \nonumber\\
&&+\frac12\sin\frac{\phi(\p) - \phi(-\p)}{2} (\Delta  d_{\p\alpha}^\dag d_{-\p\alpha}^\dag + \Delta^* d_{-\p\alpha} d_{\p\alpha})\nonumber\\
&&+  \cos \frac{\phi(\p) - \phi(-\p)}{2} (\Delta  d_{\p\up}^\dag d_{-\p\dn}^\dag + \Delta^* d_{-\p\dn} d_{\p\up}).
\label{eq:Hlow}
\eeqa
The first two lines of Eq. (\ref{eq:Hlow}) represent two decoupled superconducting bands with the spin enslaved to the local $\h(\p)$ direction. For $|\Delta| \ll \max(J,|h|)$, the superconducting interband coupling [the third line of Eq. (\ref{eq:Hlow})] provides only a small perturbation, which can be neglected. Then, within a given band, the superconductivity  corresponds to an equal spin pairing with an effective p-wave order parameter amplitude $ h_z(\p)\Delta/[2|\h(\p)|]$. The transformation from s-wave in the original model, to p-wave upon projection, is an inevitable consequence of superconductivity needing to conform to the lifted spin degeneracy. The same transformation can be applied to other combinations of magnetic fields, exchange interactions, or spin-orbital effects. The  effective order parameter  will depend on the details: for instance, for 2D Rashba superconductors, the induced effective superconductivity is  a chiral p$_x\pm i$p$_y$ due to the different $\h(\p)$ configuration around the Fermi surfaces \cite{noteR, GR}. It is important that the effective exchange field acting on electrons, $\h$, be momentum-dependent. Otherwise, s-wave superconductivity projected onto either spin-up or spin-down Fermi surface would give zero, as is the case for a uniform ferromagnet, $\h = (J,0,0)$ \cite{piet}. For HMS $h_z(\p)$ vanishes when $p_z=0$ and a line of nodes in the superconducting gap appears at the Fermi surface \cite{bulaevskii-kulic, buzdin-kulic}. 

\section{Origin of surface Zero modes}

In the idealized  1D case (or for any fixed pair $p_x$ and $p_y$) the Hamiltonian (\ref{eq:H2}) is unitarily equivalent to the model of 1D Rashba superconductor \cite{das Sarma, von Oppen, been2, loss}. Since in a range of parameters this model gives isolated Majorana edge modes,  from this mapping it immediately follows that for certain parameter range there will be Majorana modes on the surface of HMS. However, a more general argument  for the existence of the Majorana modes follows from a direct solution of the Bogoliubov-de Gennes (BdG) equations for spinless fermions in the experimentally important regime of weak superconductivity.  For convenience, we  keep the surface perpendicular to the $\z$ axis, but take an arbitrary non-chiral p-wave order parameter, $\Delta(\p)$. If the line of fixed $(p_x,p_y)$ crosses the Fermi surface in two points, $\p_R=(p_x,p_y, p_{zR})$ and $\p_L = (p_x,p_y, p_{zL})$, the quasiclassical BdG equations in the vicinity of these points read \cite{deGennes}
\beqa
-i \nu_{R} \partial_z u_R  + \Delta_R v_R = E u_R\\
i \nu_{R} \partial_z v_R  + \Delta_R u_R = E v_R
\eeqa
for ``right-movers" with velocity $\nu_R = \partial\ve/\partial p_z|_{\p_R}$, and similarly for the ``left-movers" (the detailed derivation is in the Appendix \ref{app:bdg}).  If the sign of $\nu_{R} \Delta_R$ is the same as of $\nu_{L} \Delta_L$, e.g., positive, then one can construct a zero energy solution normalizable for $z > 0$ that vanishes at $z =0$ as
\beq
\left(\begin{array}c u\\ v\end{array}\right) = \left(\begin{array}c 1\\ i\end{array}\right) \left(e^{(ip_{zR} - \frac{\Delta_R}{\nu_R})z } - e^{(ip_{zL} - \frac{\Delta_L}{\nu_L})z}\right) e^{i p_xx + i p_yy}.
\eeq
Hermitian conjugation relates the zero modes at $(p_x,p_y)$ and $(-p_x,-p_y)$. Therefore, their linear superpositions are the canonical Majorana zero modes (see Appendix \ref{app:zm} for details).
If, on the other hand, $\nu_R\Delta_R \nu_L\Delta_L <0$, no normalizable solution that satisfies boundary conditions can be constructed at $E=0$ and thus there are no zero energy Majorana modes. In addition, if there is an even number of pairs of points where the line of fixed $(p_x,p_y)$ crosses the Fermi surface, then, generically, instead of  Majorana modes there will be complex fermions at finite energies.

\begin{figure}
\includegraphics[width = 3.5 in, height = 2.2 in]{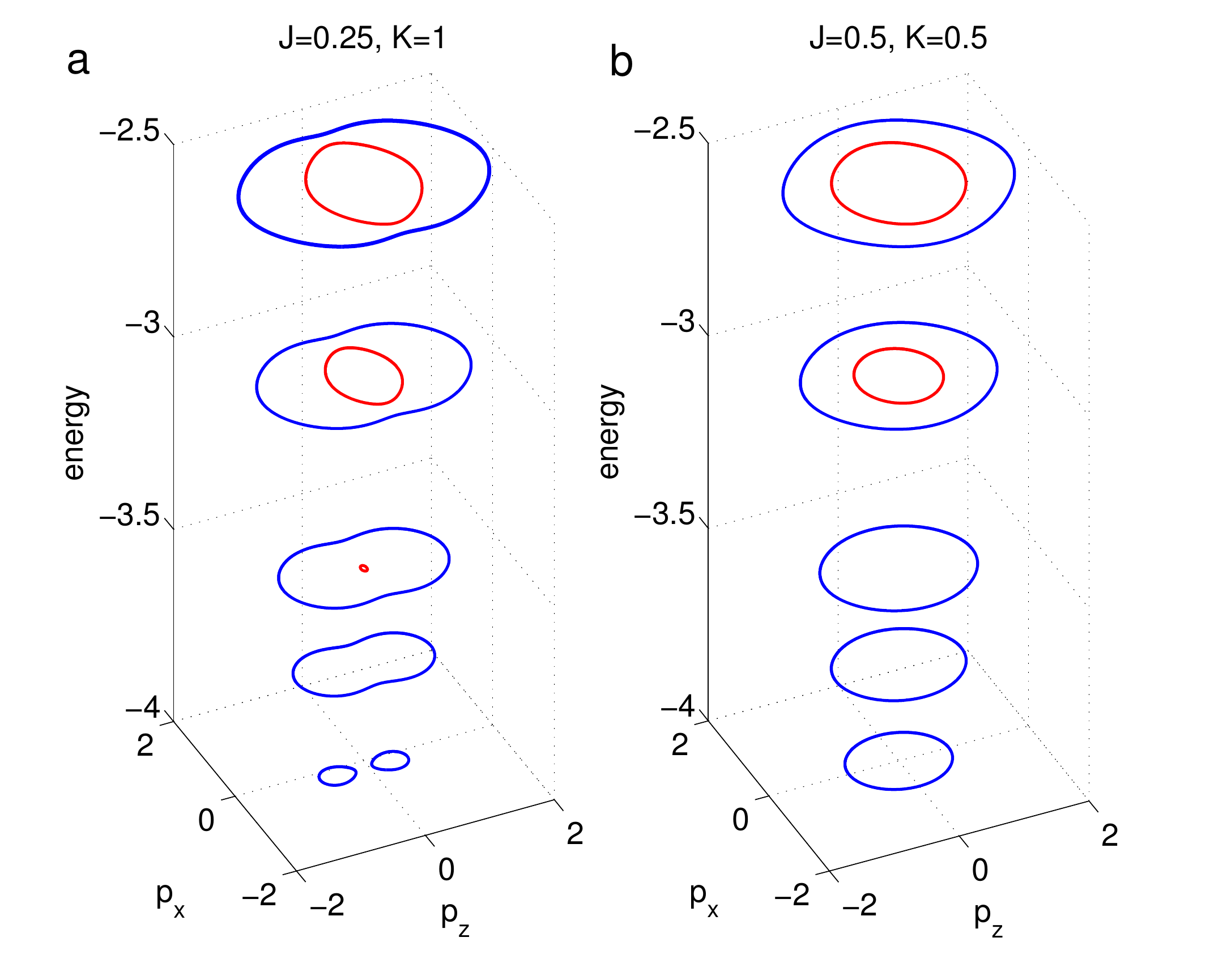}
\includegraphics[width = 1.6 in]{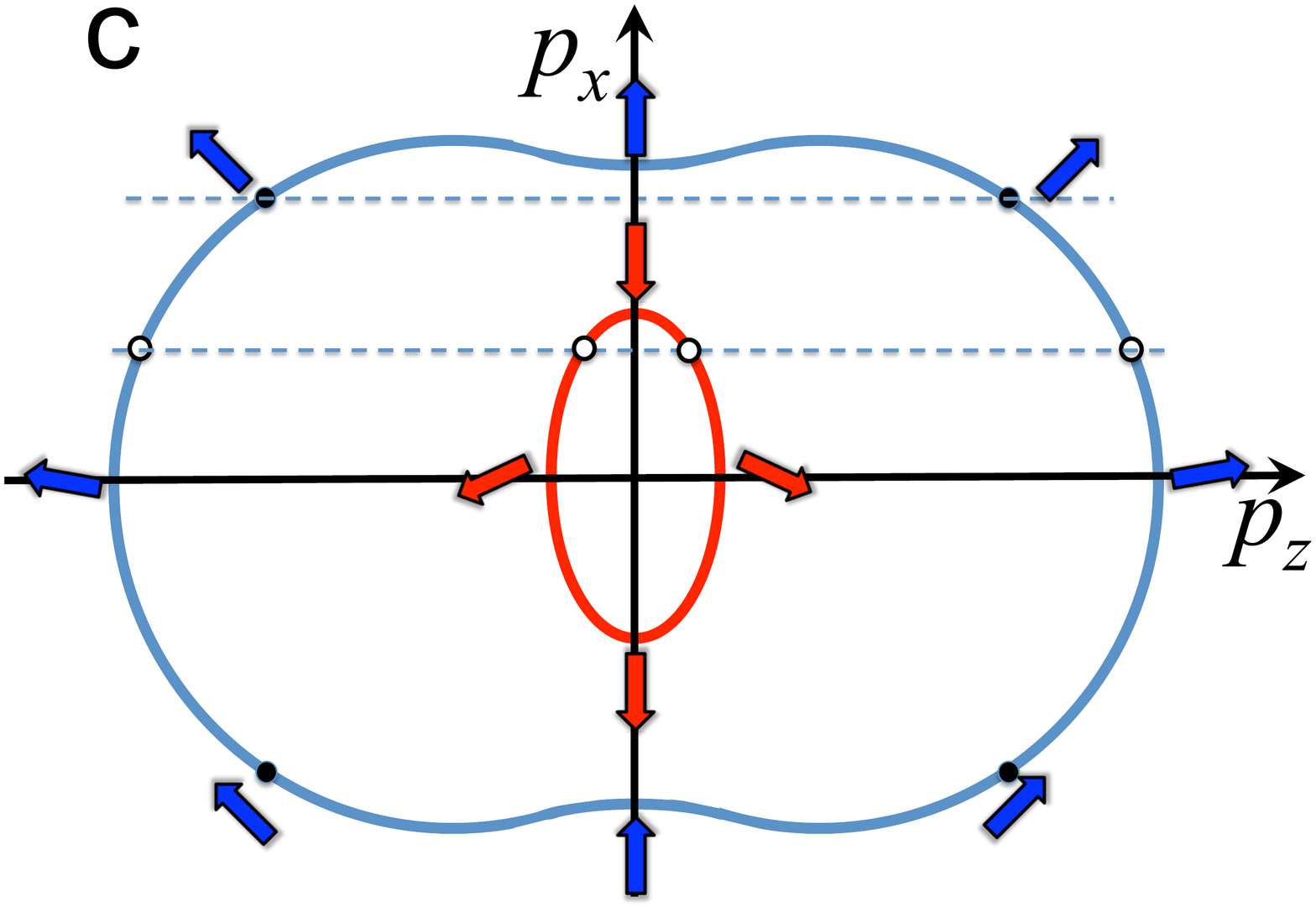}
\includegraphics[width = 1.6 in]{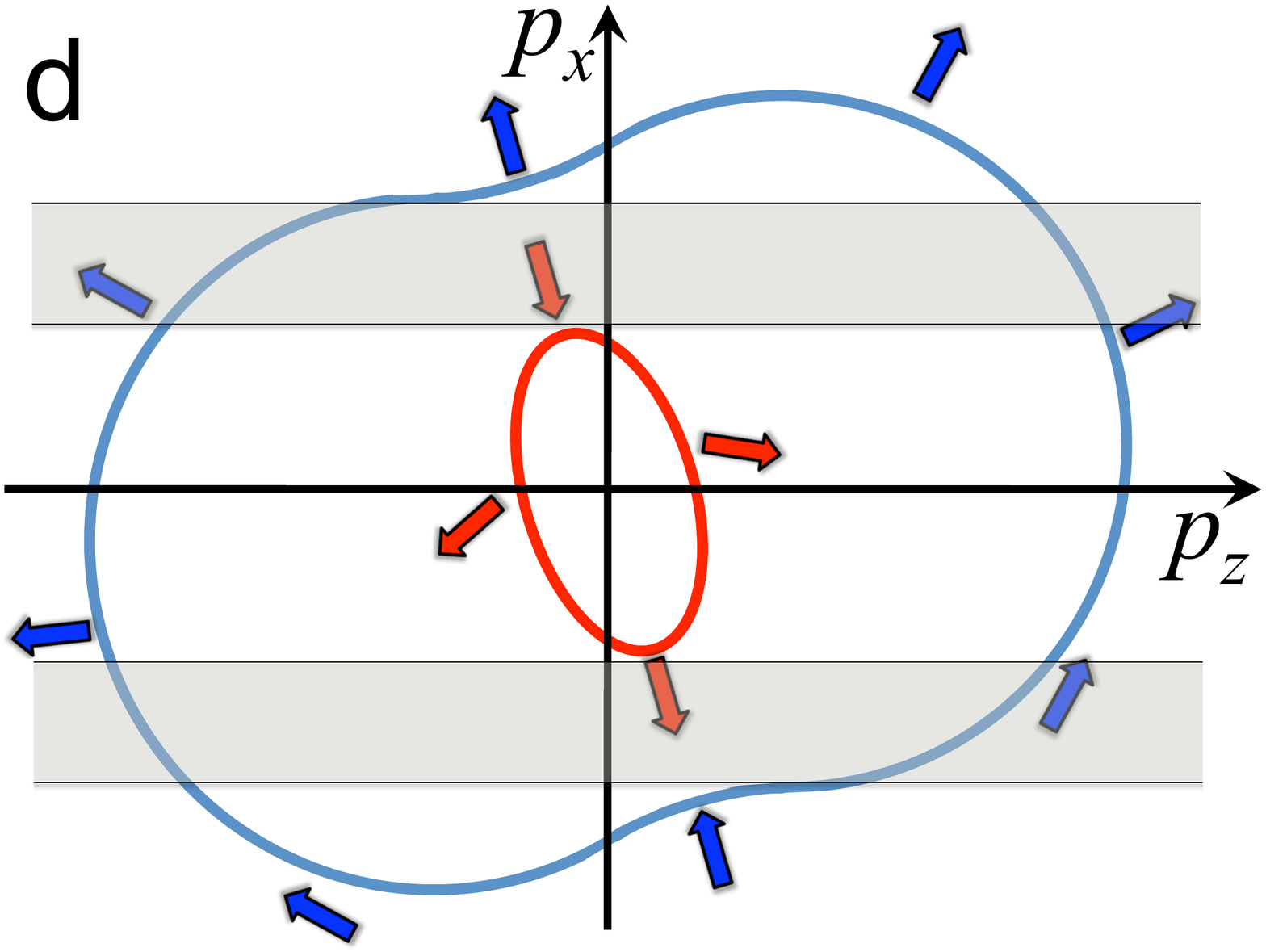}
\caption{(Color online) 2D cuts at $p_y =0$ through the Fermi surfaces of a helical magnet at different energies: (a) for the case $J < K/2$ and (b) for $J > K/2$. The outer/inner (blue/red) line corresponds to the lower/upper energy band. (c) The arrows indicate the spin polarizations of momentum eigenstates on the Fermi surfaces. The dashed lines parallel to the $p_z$ axis can have either two (upper line) or four crossings (lower line) with the Fermi surfaces. The former leads to undoubled Majorana surface modes, while the latter does not. Panel (d) illustrates the case of arbitrary orientation of the helical axis, and thus also of the p-wave superconductivity, relative to the surface normal ($\z$). The undoubled Majorana modes exist in the shaded interval of momenta parallel to the surface: there, both the order parameter and the $\z$-component of Fermi velocity change sign upon reflection from the interface.}\label{fig:FS}
\end{figure}

The situation is illustrated in Fig. \ref{fig:FS}, where we used the band structure of a helical magnet on a cubic lattice with the nearest neighbor hopping of unit strength and lattice constant taken as a unit of length [$\ve (\p) = -2 (\cos p_x + \cos p_y + \cos p_z)$]. In panels (a) and (b) we show a series of energy-dependent $p_y = 0$ cuts of the Fermi surfaces for the cases $J < K/2$ and $J > K/2$, respectively (the Fermi surface topologies at low energies differ in these cases). In Fig. \ref{fig:FS}c, the dashed lines indicate the lines of constant $p_x$ and $p_y$, which can cross either one or both Fermi surfaces. The case shown by the upper dashed line satisfies the conditions for existence of undoubled Majoranas: there are only two crossing points and between them, both velocities in the $\z$ direction, and the $p_z$-wave order parameter, change sign. On the other hand, for the lower dashed line, there are crossings on both Fermi surfaces. In this case, any finite interband coupling will lead to fusion of the two Majorana modes belonging to the individual bands into a finite energy fermion. In Fig. \ref{fig:FS}d we illustrate a more general case of alignment of the axis of the non-chiral p-wave superconducting order parameter relative to the surface. In this case, the Majorana modes will exist within the shaded range of momenta $p_x,p_y$. Interestingly, such finite concentration of Majorana zero modes has also been predicted to exist in some time-reversal-invariant systems \cite{TRI}: the results presented here indicate that the presence or absence of time-reversal invariance is not the determining factor for the presence or absence of a large density of zero-energy surface states.

\section{Numerical results}

To verify the validity of the above considerations, we have performed numerical simulations of the HMS Eq. (\ref{eq:H}) in two dimensions on a square lattice.  We consider the system with a fixed number of sites in the $\z$ direction and with translational invariance  in the $\x$ direction.  The helical direction $\K$ is chosen either along $\z$ or $\x$ (the results for other orientations of $\K$ are presented in Appendix \ref{app:num}).    From the above argument based on the change of sign of velocity and order parameter upon reflection, we expect to obtain Majorana modes in the $\K||\z$ but not in the  $\K\perp\z$ case. It is convenient to define $p_\pm = \cos^{-1}(\cos K/2 + \mu \pm J/2 )$. For weak superconductivity, $|\Delta| << |J|$, when both bands are populated, $p_\pm$ are real and the Majorana fermions exist in the interval $(-|p_+|, -|p_-|)  \bigcup (|p_+|, |p_-|)$. When only one band is populated and $p_+$ is real (the Fermi surface is simply connected) then the Majorana modes exist for $p_x$ in the range $(-p_+, p_-)$. 

\begin{figure}
\includegraphics[width = 3.7 in, height = 3. in]{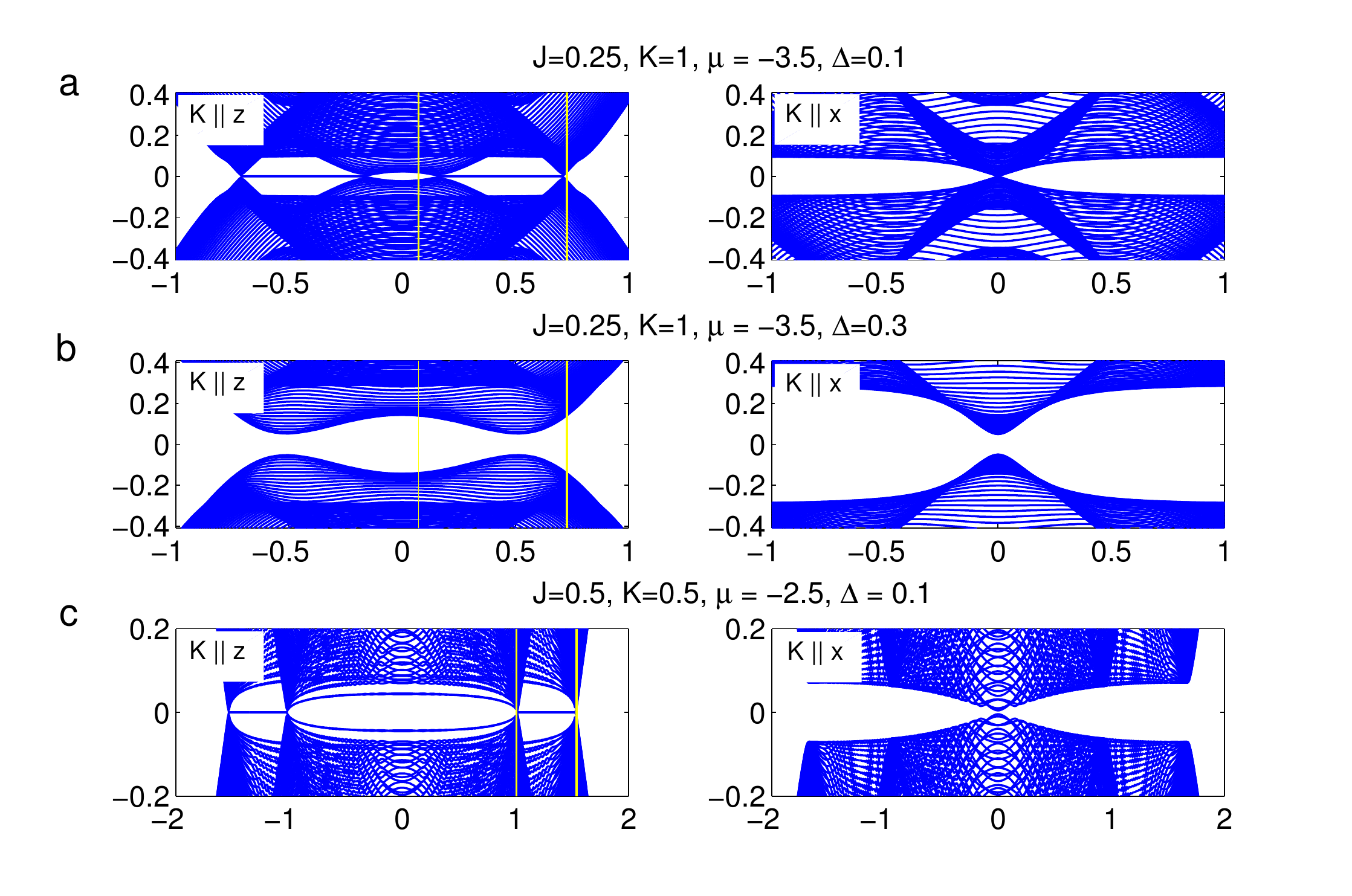}
\caption{(Color online) Results of numerical simulations of model Eq. (\ref{eq:H}) on a square lattice with nearest neighbor hopping.  The system is periodic in the $\x$ direction and has open  boundaries in the $\z$ direction. The size of the system in the $\z$ direction is 150 sites.    The helical wavevector $K$ is either along $\z$ (left column) or along $\x$ (right column). The horizontal axes are the momentum along the edge (the lattice constant is the unit of length); the vertical axes are the energy in units of intersite hopping.
Case (a) corresponds to the intermediate strength superconductivity, $\Delta/J = 0.4$ ($J,K$ parameters from Fig. \ref{fig:FS}a, the third pair of Fermi surfaces from the top). The range of momenta $p_x>0$ where unpaired Majorana modes are expected from the small-$\Delta$ consideration is shown by vertical yellow lines. For $K|| x$ there are no Majorana modes, as expected, since the sign of the order parameter does not change under reflection from the interface. The near-zero energy modes at $p_x = 0$ are due to the gapless superconductivity in the bulk. In case (b) the amplitude of superconducting pairing has overcome the band splitting induced by magnetic ordering. A full superconducting gap opens and there are no zero energy modes for either orientation of $K$ relative to the sample boundary. Case (c) corresponds to weak superconductivity, $\Delta/J = 0.2$, for the parameters in Fig. \ref{fig:FS}b (the top pair of Fermi surfaces).  The Majorana modes exist in the expected range of momenta $p_x$ -- marked by vertical (yellow) lines -- as discussed in the main text.
}\label{fig:num}
\end{figure}

In Fig. \ref{fig:num} we present the results for the electronic spectra for a set of parameters representing large and small $\Delta/J$ and different chemical potentials. The spectra are doubled as they include the Bogoliubov redundancy, and only the non-negative energy states should be considered. We see that in an interval of $p_x$ that approximately coincides with the interval $(p_-, p_+)$ indicated by the vertical yellow lines, there remains only one low energy mode. We verified that the energy of the mode vanishes exponentially quickly with the increasing distance between the edges and that the probability amplitude is equally split between the two edges of the sample, as expected for Majorana states. As can be seen from Fig. \ref{fig:num}a, $p_\pm$  give a good estimate of the momentum range of Majorana modes even for rather large $\Delta/J = 0.4$, and the agreement becomes excellent for $\Delta/J = 0.2$ (Fig. \ref{fig:num}c).  When the helical direction is aligned with the edge, $\K||\x$, there are no zero-energy Majorana modes. The low energy states near $p_x = 0$ are due to the gapless nature of the effective $p_x$-wave order parameter in this case. Numerical results for intermediate angles between the helical direction and the sample surface are presented in Appendix \ref{app:num}.  The flat section with zero modes appears in a finite range of angles, consistent with the general argument based on the 
BdG consideration of the previous section.
Numerically we can also address the regime of strong pairing. As expected, when pairing can overwhelm the magnetic exchange gap, the system undergoes a quantum phase transition into a fully gapped state adiabatically connected to the trivial s-wave superconductor. In this regime, naturally, there are no longer any Majorana surface states, Fig. \ref{fig:num}b.

\section{Effects of confinement}
For quantum information applications it is important to be able to isolate a small number of Majorana zero modes. This can be done by shaping sample into a wire. Suppose that a 2D wire is confined in the $\x$ direction when the helical axis is pointing along $\z$.
When confined to a cylinder of perimeter $L$, the allowed values of the momentum along the edge are $p_x = 2\pi n/L$ with $n$ any integer.  Since the modes with $n\ne 0$ come in pairs and $n = 0$ is unique, the odd number of Majorana modes at a given edge persists as long as there is only one (non-superconducting) Fermi surface. As soon as the the upper energy band starts to be populated, the number of Majorana modes becomes even.
In the case of a ribbon (the width in the $\x$ direction is much smaller than in the helical -- $\z$ -- direction), one can first construct transverse modes $\psi_{n_x} (x/W_x) e^{i p_z z}$ from the superpositions of the states with $p_x = \pm\pi n_x/W_x$, where $n_{x}$ are positive integers. These states can be easily made to satisfy the zero boundary conditions along the $\x$ axis since they have the same spin quantization axis. Since different transverse modes correspond now to different $n_x>0$, even when both high and low energy bands cross the Fermi surface the number of Majorana surface modes per edge, $N_M \approx (p_+ - p_-)W/\pi$, can be either even or odd.  In fact, with increasing chemical potential, $N_M$ decreases since the distance between the Fermi surfaces of the two bands shrinks.  Assuming parabolic electron dispersion, $N_M \approx J/ (p_F\delta p_x) $, i.e. $N_M $ is given by the ratio of the exchange coupling to the 1D single particle level spacing. Consequently, even for large total number of transverse modes $\sim p_F/\delta p_x$, $N_M$ can be $\sim 1$. For comparison, in the case of a 3D bar with finite dimensions in both $\x$ and $\y$ directions,  $N_M \sim 2m J/(\delta p_x \delta p_y)$. Even in this case, if $W_x \ne W_y$, $N_M$ will be either odd or even as a function of system parameters ($\mu, W_{x,y},  J$). The dependence of the parity of the Majorana mode number on the system parameters, such as the transverse dimensions, resembles the situation occurring in the case of Rashba superconductors, thought there the dependence is somewhat more intricate \cite{Potter, stanescu2}.

\section{summary}

We have shown that ideal surfaces of HMS host a finite density of zero energy Majorana modes.   When confined to wires, the number of Majorana modes can be reduced to a small number. The relatively large relevant energy scales -- with critical temperatures for superconductivity reaching up to 10 K and exchange interaction of about 100 K,  HMS offer attractive conditions for the investigations of the Majorana physics.

\begin{acknowledgements}
{ We thank D. Podolsky, A. Shnirman, J. Li, M. B\"uttiker, and {\O}. Fischer for discussions.  IM thanks University of Geneva, where this work was completed,  for hospitality. This
work was carried out under the auspices of the National Nuclear
Security Administration of the U.S. Department of Energy at Los
Alamos National Laboratory under Contract No. DE-AC52-06NA25396
and supported by the LANL/LDRD Program. AFM gratefully acknowledge financial supports from the NCCR MaNEP, QSIT and from the Swiss National Science Foundation. }
\end{acknowledgements}

\appendix

\section{Derivation of the Bogoliubov-de Gennes (BdG) equations for spinless fermions}\label{app:bdg}
In this section, for the sake of completeness, we derive the BdG equations (5) and (6) of the main text. For simplicity of presentation, we will assume that the Fermions have parabolic dispersion, and that the superconducting order parameter is a non-chiral p-wave with the lobes oriented along a unit vector $\hat n$. Then the Hamiltonian in the real space representation is
\beqa
H &=&\int d\rr \,\left[\psi^\dag \left(\frac{\nabla^2}{2m} - \mu\right) \psi \right.\\
&&\left.+ i\alpha\psi^\dag(\hat\partial\cdot \hat n)\psi^\dag +  i\alpha^*\psi(\hat\partial\cdot \hat n)\psi\right],
\eeqa
where $\psi$ and $\psi^\dag$ are the fermion filed annihilation and creation operators, $\{\psi^\dag(\rr) , \psi(\rr')\} = \delta({\rr - \rr'}) $, $m$ is the fermion mass, $\hat \partial = (\partial_x, \partial_y, \partial_z )$, and $\alpha$ determines the amplitude of the order parameter, which can smoothly vary in space.
The Bogoliubov quasiparticles $\gamma$ are the eigenmodes of this Hamiltonian, i.e., 
\beq
[H,\gamma^\dag] = E \gamma^\dag.\label{eq:gamma}
\eeq
They can be expressed in terms of the fermion field operators as 
\beq
\gamma = \int d\rr[u\psi + v\psi^\dag],
\eeq
with $u(\rr)$ and $v(\rr)$ being scalar functions. Performing commutation according to Eq.~(\ref{eq:gamma}) and collecting coefficients of $\psi$ and $\psi^\dag$, we obtain the following equations for $(u,v)$:
\beqa
\left(\frac{\nabla^2}{2m} - \mu\right) u - 2i\alpha ^*(\hat\partial\cdot \hat n) v = Eu,\\
-\left(\frac{\nabla^2}{2m} - \mu\right) v - 2i \alpha(\hat\partial\cdot \hat n) u = Ev.
\eeqa
In the vicinity of a particular Fermi point $\p_F$, we can factor the fast oscillating and the slow parts,
\beq
\left(\begin{array}{ c} u \\v\end{array}\right) = \left(\begin{array}{ c} \tilde u \\\tilde v\end{array}\right) e^{i\p_F \rr}, 
\eeq
and for the slow part, finally, the BdG equations are
\beqa
-i\bnu_F\cdot \hat\partial u + 2\alpha ^*(\p_F\cdot \hat n) v = Eu,\\
i \bnu_F\cdot \hat\partial v + 2 \alpha(\p_F\cdot \hat n) u = Ev.
\eeqa
The Fermi velocity, for arbitrary non-interacting dispersion $\varepsilon_\p$ is  $\bnu_F = \partial \varepsilon_\p/\partial\p $.
For non-chiral order parameter, and for the situation when the translational invariance is broken in the $\z$ direction only (e.g. by the presence of a surface), these equations are precisely Eqs.~(5) and (6) of the main text.

\section{Zero energy solutions of the BdG equations in the presence of a surface}\label{app:zm}
We are interested in the existence of the zero energy quasiparticle states localized near the $z = 0$ surface. We will attempt to construct these solutions as a combination of incoming and outgoing plane waves that correspond to the same values of $p_x$ and $p_y$,  in the vicinity of either $\p_R=(p_x,p_y, p_{zR})$ or $\p_L = (p_x,p_y, p_{zL})$ Fermi points.
Near a particular Fermi point, we are looking for the solution of 
\beqa
-i\nu\partial_z u + \Delta v = 0,\\
i \nu\partial_z v + \Delta u = 0.
\eeqa
(Here we dropped the irrelevant global phase of the order parameter $\Delta$ and tilde over $(u,v)$). These equations can be easily integrated to find that 
\beq
u\pm iv = A_\pm e^{\mp \Delta\, z/\nu}.
\eeq
The solution has to be normalizable at $z \ge 0$.  Therefore, for a given sign of $\Delta/\nu$, e.g. positive, $A_+ = 0$, and consequently $u = -iv$.  
Including now also the fast oscillating part, the individual solutions near each of the Fermi points are proportional to
\beq
\left(\begin{array}{ c} 1\\ i\end{array}\right) e^{i\p_F \rr - \Delta\, z/\nu}.
\eeq
Note that as long as the sign of  $\Delta/\nu_F$ is the same for both Fermi points, the spinor part of the solutions is the same. We can therefore take a linear combinations of them such that the full wave function vanishes at the boudnary, $z = 0$.  The corresponding Bogoliubov quasiparticle is
\begin{widetext}
\beq
\gamma = \int d\rr[\psi(\rr) + i\,\psi^\dag(\rr)]\left(e^{(ip_{zR} - \Delta_R/\nu_R)z} - e^{(ip_{zR} - \Delta_R/\nu_R)z}\right) e^{ip_x x + i p_y y}.
\eeq
\end{widetext}
This is an operator that creates a zero energy quasiparticle near an infinite surface for a given value of $(p_x,p_y)$.  Despite being a non-degenrate zero-energy solution, this state does not completely satisfy the criteria for Majorana zero mode, i.e., $\gamma\ne \gamma^\dag$. However, for a p-wave superconductor with a electronic dispersion that is symmetric with respect to $\p\to -\p$, for every solution $\gamma$ at a given $(p_x,p_y)$, there is a corresponding solution $\gamma'$ at $(-p_x,-p_y)$.  They form a complex conjugate pair, $\gamma^\dag = \gamma'$. The ``canonical" Majorana zero modes can be constructed out of them as $\gamma + \gamma'$ and $i(\gamma - \gamma')$.  For an ideal surface, there is no mixing between different $(p_x,p_y)$ states and consequently for each allowed $(p_x,p_y)$ point there is only half of a complex fermion mode.

\section{Numerical results for a general orientation of the axis of p-wave order parameter relative to the sample surface}\label{app:num}
\begin{figure}
\includegraphics[width=1.1\columnwidth]{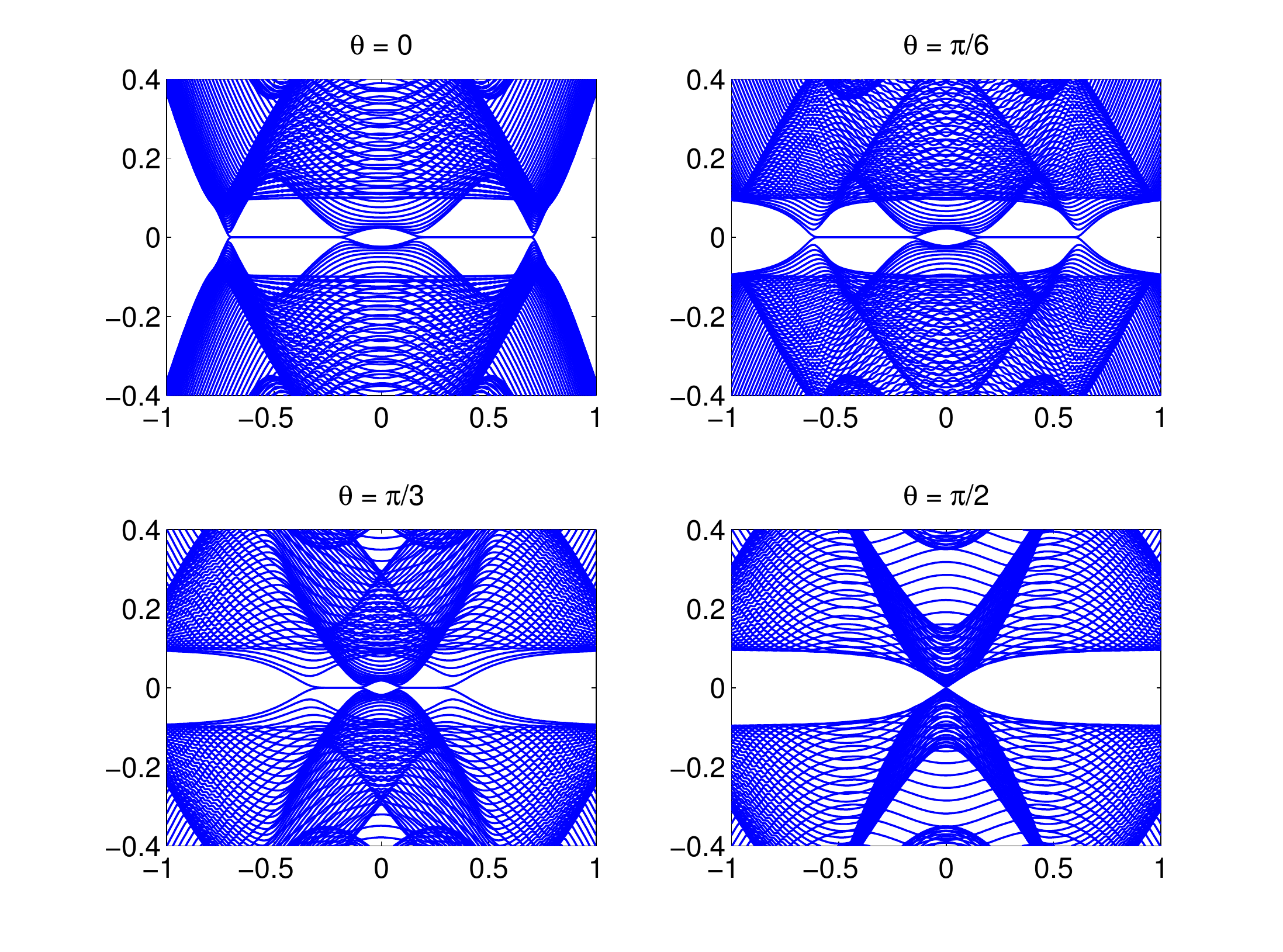}
\caption{(Color online) Results of numerical simulations of model Eq. (1) of the main text on a square lattice with nearest neighbor hopping.  The parameters are the same as in Fig. 2a, $J=0.25,\ K=1,\ \mu = -3.5,\ \Delta=0.1$. The system is periodic in the $\x$ direction and has open  boundaries in the $\z$ direction. The horizontal axes are the momentum along the edge (the lattice constant is the unit of length); the vertical axes are the energy in units of intersite hopping. The size of the system in the $\z$ direction is 150 sites.    Different plots correspond to various angles $\theta$ between the helical wavevector $K$ and the $\z$-axis ($K$ is assumed to be lying in the $\x$-$\z$ plane). 
The horizontal axes are the momentum along the edge (the lattice constant is the unit of length); the vertical axes are the energy in units of intersite hopping.
Plots for $\theta = 0$ and $\theta = \pi/2$ are the same as in  Fig. 2a.  
For intermediate angles we see that the range of momenta where zero-energy Majorana modes are present, changes gradually.
}\label{Fig:theta}
\end{figure}

In Fig. \ref{Fig:theta} we demonstrate that the zero energy Majorana modes exist for a wide range of the relative orientations of the surface normal and the order parameter axis, as has been argued in the main text based on the analysis of the BdG equations. This proves that the Majorana modes generically appear on the surfaces of non-chiral p-wave superconductors, as long as the translational invariance in the plane of the surface is preserved.

\begin{thebibliography}{10}
\bibitem{noteR} {The effective p-wave superconductivity in the case of HMS is distinct from the 2D Rashba superconductor in an important
respect. Since in HMS the rotation to the local spin quantization axis corresponds to a non-singular gauge transformation, the resulting superconducting order parameter can be chosen real. In the case of Rashba spin-orbit, transforming to the local spin quantization reference frame requires a singular gauge transformation, which would  lead to an effectively chiral $p_x \pm ip_y$ order parameter \cite{Fu-Kane}.}
\end{thebibliography}
\end{document}